\documentclass{aa}  

\usepackage{graphicx}
\usepackage{txfonts}

\begin{document} 

  \title{Suzaku study of gas properties along filaments of A2744 }

   \author{Y. Ibaraki \inst{1} 
          \and
          N. Ota \inst{1} \fnmsep\thanks{\email{naomi@cc.nara-wu.ac.jp}}
          \and 
          H. Akamatsu \inst{2}
          \and 
          Y.-Y. Zhang\inst{3}         
          \and
          A. Finoguenov\inst{4}
           }

   \institute{Department of Physics, Nara Women's University, Kitauoyanishimachi, Nara, Nara 630-8506, Japan
         \and
         SRON Netherlands Institute for Space Research, Sorbonnelaan 2, 3584 CA Utrecht, The Netherlands
         \and
	Argelander Institute for Astronomy, Bonn University, Auf dem H\"{u}gel 71, 53121 Bonn, Germany
	\and
	Department of Physics, University of Helsinki, Gustaf H\"{a}llstr\"{o}min katu 2a, FI-00014 Helsinki, Finland
             }

   \date{Received --; accepted --}

 
  \abstract
  {We present the results of {\it Suzaku} observations of a massive galaxy
    cluster \object{A2744}, which is an active merger at $z=0.308$.}
  {By using long X-ray observations of \object{A2744}, we aim to
    understand the growth of the cluster and the gas heating process
    through mass accretion along the surrounding filaments.}
  {We analyzed data from two-pointed {\it Suzaku} observations of
    \object{A2744} to derive the temperature distribution out to the
    virial radius in three different directions.  We also performed a
    deprojection analysis to study radial profiles of gas temperature,
    density, and entropy and compared the X-ray results with
    multi-wavelength data to investigate correlations with the surface
    density of galaxies and with radio relics. }
  {The gas temperature was measured out to the virial radius $r_{200}$
    in the north-east region and to about $1.5r_{200}$ in the
    north-west and south regions. The radial profile of the gas
    temperature is rather flat and the temperature is very high (even
    near $r_{200}$); it is comparable to the mean temperature of this
    cluster ($kT=9$~keV). These characteristics have not been reported
    in any other cluster. We find a hint of temperature jump in the
    northeast region whose location coincides with a large radio
    relic, indicating that the cluster experienced gas heating because
    of merger or mass accretion onto the main cluster. The temperature
    distribution is anisotropic and shows no clear positive
    correlation with the galaxy density, which suggests an
    inhomogeneous mass structure and a complex merger history in
    \object{A2744}. }
   {}


   \maketitle
%

\section{Introduction}\label{sec:intro}
\object{A2744} at $z=0.308$ (\object{AC118} or
\object{RXC~J0014.3--3022}) is one of the most actively merging
clusters, nicknamed ``Pandora's cluster''
\citep[e.g.,][]{2011ApJ...728...27O}. This cluster is also known as a
gravitational lens, with an enormous mass
\citep[$M=1.8\times10^{15}~M_{\odot}$ for $r <1.3~{\rm
  Mpc}$;][]{2011MNRAS.417..333M}. \citet{2007A&A...470..425B}
discovered the existence of two large-scale filamentary structures
that extend beyond the virial radius of A2744. Given this well-defined
large-scale structure, A2744 offers a unique opportunity to study the
formation and growth of a massive cluster in the distant universe
through merging and mass accretion along the filaments.

The mass structure and characteristics of the intracluster medium
(ICM) in the core of A2744 have been extensively studied at various
wavelengths. In the X-ray region, the cluster exhibits two emission
peaks that correspond to the main cluster and to a subcluster in the
northwest. A bow shock was discovered in the subcluster by {\it
  Chandra} \citep{2004MNRAS.349..385K}. Consistent with this
discovery, in the central region of the system, two over-densities of
member galaxies appear that are clearly separated from the two X-ray
sub-components visible in the {\it XMM-Newton} image
\citep[e.g.,][]{2008A&A...483..727P,2004A&A...413...49Z,2006A&A...456...55Z,2005A&A...442..827F}.
\citet{2011MNRAS.417..333M} performed a detailed lensing and X-ray
analysis that suggests the southern and northwestern cores are
post-merger and exhibit a morphology similar to that of the Bullet
Cluster viewed from a different angle.

The radio halo in A2744 is one of the most luminous and the bulk of
the diffuse radio emission is centered on the main cluster. The
cluster also hosts a large radio relic at a projected distance of
2~Mpc \citep{2001A&A...369..441G,2001A&A...376..803G}. This suggests
that shock heating is at work; however, due to the lack of deep X-ray
data, the detailed characteristics of the ICM at the location of relic
have yet to be clarified. Furthermore, within $r=1$~Mpc the dynamical
mass is calculated to be $(1.4-2.4)\times10^{15}~{\rm M_{\odot}}$
\citep{2006A&A...449..461B}. This value is greater than the X-ray
hydrostatic mass by $\sim40$\%, and it is likely to be overestimated
because of mergers.  Thus, A2744 calls for an assessment of the
characteristics of its complex ICM beyond the virial radius.

The X-ray Imaging Spectrometer \citep[XIS;][]{2007PASJ...59S..23K}
onboard the {\it Suzaku} satellite \citep{2007PASJ...59S...1M} enabled
observations of diffuse ICM emission up to their virial radius because
of the low and stable background.  Thus far, XIS has measured
temperature profiles in the cluster outskirts for about twenty nearby
objects, all of which show a temperature drop by a factor of two to
four at large radii \citep[for a review,
see][]{2013SSRv..tmp...56R}. Recently, \citet{2013ApJ...766...90I}
reported a similar temperature decrease in a cool-core cluster A1835
at $z=0.25$. Beyond that redshift, we still lack data. Theoretically,
the mass accretion and formation shock may preferentially occur at
high redshifts. Thus, a detailed study of the outer regions of distant
mergers is important to probe the formation history of clusters.

With the goal of understanding cluster growth via mass accretion and
gas heating, we focus on A2744 and measure the large-scale temperature
distribution with {\it Suzaku}.  This paper is organized as follows:
in \S\ref{sec:obs} the {\it Suzaku} observations and data reduction
are presented, in \S\ref{sec:analysis} the spectral analysis and
results are shown, in \S\ref{sec:discussion} the temperature profile
of A2744 is compared with other clusters and the ICM characteristics
are discussed based on a comparison with the galaxy density in the
filaments and the radio relic, and in \S\ref{sec:summary} we summarize
the results.

Throughout the paper, the cosmological model is adopted with matter
density $\Omega_{M}=0.27$, cosmological constant
$\Omega_{\Lambda}=0.73$, and Hubble constant $H_0=70~{\rm
  km\,s^{-1}\,Mpc^{-1}}$.  At the cluster redshift ($z=0.308$),
$1\arcmin$ corresponds to 274~kpc.  Unless otherwise noted, the
specified errors indicate a 90\% confidence interval.

\section{Observation and data reduction}\label{sec:obs}

\begin{table*}[htb]
\caption{{\it Suzaku} observation log}\label{tb:x_data}
\centering
\begin{tabular}{llllll} \hline\hline
  Object & OBSID & Date & RA$^{\mathrm{a}}$& DEC$^{\mathrm{a}}$& Exposure~(sec)$^{\mathrm{b}}$\\ \hline 
  A2744 Center &802033010& 2007 Jul 19--23 & 00:14:9.5 & $-30$:20:40.6 & 150583 \\ 
  A2744 South &805015010& 2010 Dec 10--12 & 00:14:3.2 & $-30$:33:2.9 & 67428 \\ \hline
\end{tabular}
\begin{list}{}{}
\item[$^{\mathrm{a}}$] Pointing coordinates in J2000.
\item[$^{\mathrm{b}}$] Net exposure time after data filtering.
\end{list}
\end{table*}

The present study uses {\it Suzaku} XIS data of A2744 taken during two
pointed observations, one centered on the main cluster and the other
on the south region. The net exposure times were 150~ks and 67~ks for
the center and the southern regions, respectively. The summary of the
observation log is given in Table~\ref{tb:x_data}. The XIS detectors
consist of four X-ray-sensitive CCD cameras, three front-illuminated
CCDs (XIS-0, XIS-2, XIS-3), and one back-illuminated CCD (XIS-1)
\citep{2007PASJ...59S..23K}. Among them, XIS-0, XIS-1, and XIS-3 were
operated in normal modes during the observations and the Space Charge
Injection option was applied \citep{2008PASJ...60S..35U}.
 
 \begin{figure}[htb]
\centering
\rotatebox{0}{\scalebox{0.4}{\includegraphics{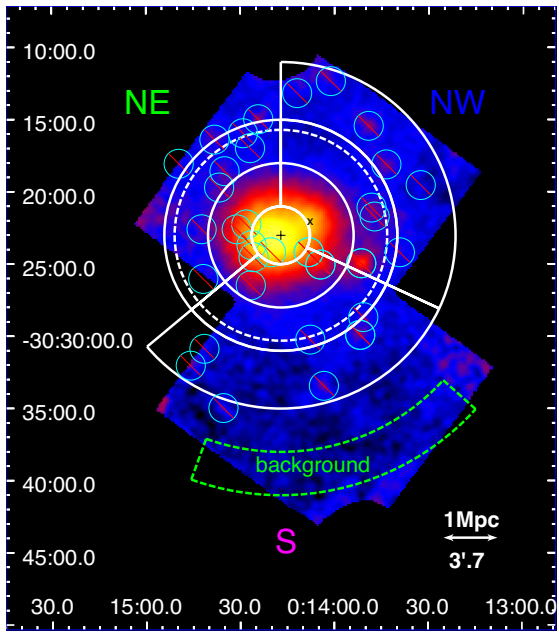}}}
\caption{{\it Suzaku} XIS-0 + XIS-1 + XIS-3 mosaic image of
  \object{A2744} in the 0.5--8~keV band. The image is corrected for
  exposure map and vignetting effects and is smoothed by a Gaussian
  function with $\sigma=24\arcsec$. No background was subtracted. 
    The X-ray peaks of the main cluster and the NW subcluster are
    marked with ``+'' and ``x'', respectively. The spectral
  integration regions ($0\arcmin-2\arcmin, 2\arcmin-5\arcmin,
  5\arcmin-8\arcmin, 8\arcmin-12\arcmin$) are shown with the white
  annuli, while point sources detected by {\it XMM-Newton} PN and MOS
  are eliminated with an $r=1\arcmin$ radius (the cyan circles with
  red diagonal lines).  The white dashed line denotes the virial
  radius $r_{200}=7\arcmin.3$. A fan-shaped $r=15\arcmin-18\arcmin$
  region in the south (green) was used to estimate background. The two
  corners of the CCD chip illuminated by $^{55}$Fe calibration sources
  are excluded from the image. }\label{fig:A2744_xis}
\end{figure}

Fig.~\ref{fig:A2744_xis} shows the XIS mosaic image of A2744 in the
0.5--8~keV energy band.  Thus, the XIS field of view covers the ICM
emission out to the cluster virial radius $r_{200}=2.0$~Mpc (see
\S\ref{subsec:global_spec}) and beyond. As seen in
Fig.~\ref{fig:A2744_xis}, the X-ray morphology is irregular, with
elongation along the west-east direction and bimodal X-ray-emission
peaks at the center.

We used cleaned event files created by the pipeline processing version
2.4 and performed the data analysis using HEASoft ver 6.12 and CALDB
version 2012-09-02 for XIS and version 2011-06-30 for the X-ray
telescopes \citep[XRT;][]{2007PASJ...59S...9S}. The XIS data were
filtered according to the following criteria: Earth elevation angle $>
10^{\circ}$, day-Earth elevation angle $> 20^{\circ}$, and the
satellite was outside the South Atlantic Anomaly.

To measure the average ICM temperature, the X-ray spectra were
extracted from the full XIS field of view in the central pointing.
Note that, throughout the spectral analysis point sources detected by
{\it XMM-Newton} were removed with an $r=1'$ circle. To investigate
the radial distributions, we analyzed the spectra integrated from
annular regions centered on the X-ray centroid, (RA, Dec)=(00:14:17.1,
$-30$:23:02.6) \citep{2004A&A...428..757O}. The radial ranges of the
spectra were $0'-2', 2'-5', 5'-8'$, and $8'-12'$ and they were divided
into the three azimuthal directions NW, NE and S
(Fig.~\ref{fig:A2744_xis}) because two filaments were identified
  in the NW and S directions \citep{2007A&A...470..425B}. The spectra
were rebinned so that each spectral bin contained more than 60~counts
for spectra in the center and 80~counts for spectra in the south.

\begin{table*}[htb]
\caption{Spectral fits to background data for CXB and Galactic components.}\label{tab:bgd}
\centering
\begin{tabular}{lllllll} \hline\hline
$\Gamma$ & $Norm$ $^{\mathrm{a}}$ & $kT_{\rm MWH}$ & $Norm_{\rm MWH}$$^{\mathrm{b}}$ & $kT_{\rm LHB}$ & $Norm_{\rm LHB}$$^{\mathrm{b}}$ & $\chi^2$/d.o.f \\
         &($\times10^{-4}$)& [keV] & ($\times10^{-3}$) & [keV] &($\times 10^{-2}$)&     \\ \hline
1.40(fixed) &  $5.83^{+0.38}_{-0.46}$ &  $0.34^{+0.24}_{-0.34}$  
& $0.38^{+2.34}_{-0.38}$ &  $0.13^{+0.47}_{-0.13}$ &  $0.14^{+1.07}_{-0.14}$  &38.7/38 \\ \hline
\end{tabular}
\begin{list}{}{}
\item[$^{\mathrm{a}}$] Photon flux of power-law model, in units of ${\rm
    photons\,keV^{-1}\,cm^{-2}\,s^{-1}}$ at 1~keV.
\item[$^{\mathrm{b}}$] Normalization of the APEC model, $Norm = \int
  n_e n_H dV/(4\pi (1+z)^2 D_A^2)~[10^{-14}{\rm cm^{-5}}]$. $D_A$ is
  the angular diameter distance to the source. An $r=20'$ uniform sky is assumed.
\end{list}
\end{table*}

The non-X-ray background was subtracted using {\tt xisnxbgen}
\citep{2008PASJ...60S..11T}. The other background components, i.e.,
the cosmic X-ray background (CXB) and the Galactic X-ray emission
arising from the Local Hot Bubble (LHB) and the Milky Way Halo (MWH)
were assessed by using the same model as used in
\citet{2011PASJ...63S.979S} and in \citet{2013A&A...556A..21O}. The
background spectra in the south ($15'<r<18'$) were fit to the model
``apec$_{\rm LHB}$+phabs(apec$_{\rm MWH}$+power-law$_{\rm CXB}$)'',
which yielded the parameters given in Table~\ref{tab:bgd}, where the
spectral normalization factors were derived assuming that these X-ray
background components have uniform surface brightness distributions
within $r=20\arcmin$ centered at the XIS optical axis.  According to
this model, the X-ray background in each spectral region was simulated
by using the {\tt XSPEC} fakeit command and then subtracted from the
observed spectrum.

We used {\tt xisrmfgen} to generate the energy response file. To take
into account the XRT's vignetting effect and a decrease in the
low-energy efficiency because of the contaminating material on the
optical blocking filter of the XIS the auxiliary response files were
calculated by using {\tt xissimarfgen} \citep{2007PASJ...59S.113I}.
For input cluster surface brightness, we assumed a $\beta$-model image
with $\beta=0.96$ and $r_{c}=2\arcmin.2$ \citep{2004A&A...428..757O}.

\section{Analysis and results}\label{sec:analysis}
\subsection{Global spectra in center}\label{subsec:global_spec}
\begin{figure}[htbp]
\centering
\rotatebox{0}{\scalebox{0.33}{\includegraphics{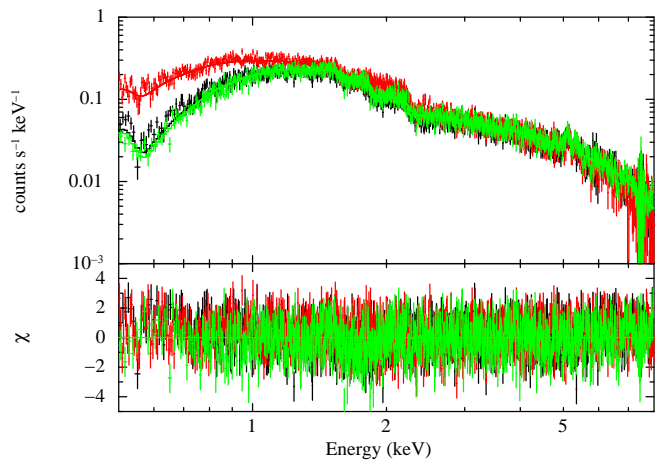}}}
\caption{XIS spectra of A2744 center.  XIS-0 (black), XIS-1 (red),
  XIS-3 (green) are shown separately.  The solid lines in the upper
  panel show the best-fit APEC model convolved with the telescope and
  detector response functions.}
   \label{fig:global_spec}%
 \end{figure}

To measure the average temperature of \object{A2744}, we first
analyzed the XIS spectra extracted from an $18'\times18'$ square in
the central pointing (Fig.~\ref{fig:A2744_xis}).  The observed
0.5--8~keV spectra of three sensors (XIS-0, XIS-1, and XIS-3) were
simultaneously fit to the APEC thermal plasma model
\citep{2001ApJ...556L..91S} (Fig.~\ref{fig:global_spec}). The cluster
redshift and Galactic hydrogen column density were fixed at $z=0.308$
and $N_{\rm H}=1.39\times 10^{20}~{\rm cm^{-2}}$ \citep[LAB
survey;][]{2005A&A...440..775K}, respectively. For metal abundance, we
used the tables in \cite{1989GeCoA..53..197A}.  The $\chi^2$ fitting
was performed using {\tt XSPEC} version 12.7.1

 The model parameters are given in Table~\ref{tab:global_spec}. The
 resulting global temperature and metal abundance are $kT = 9.0 \pm
 0.2$~keV and $Z=0.21 \pm 0.02$~solar, respectively, which agree well
 with the previous {\it ASCA} measurement
 \citep{2004A&A...428..757O}. Substituting a mean temperature of
 $kT=9.0$~keV into the $R-T$ relationship \citep{2005A&A...441..893A}
 gives a virial radius of $r_{200}=2.0~{\rm Mpc}$ or 7\arcmin.3.

\begin{table}
\caption{APEC model parameters for global spectra of A2744}\label{tab:global_spec}
\centering
\begin{tabular}{llll} \hline\hline
$kT$~[keV] & $Z$~[solar] & Norm & $\chi^2$/d.o.f. \\ \hline
$9.00^{+0.20}_{-0.20}$ & $0.21^{+0.02}_{-0.02}$ & $9.91^{+0.08}_{-0.08}\times10^{-3}$ & 2993/2739 \\ \hline
\end{tabular}
\end{table}

\subsection{Annular spectra in three directions}\label{subsec:annular_spec}
To derive the radial temperature profiles along the filaments, we
extracted spectra from the concentric rings: $0'-2'$, $2'-5'$,
$5'-8'$, and $8'-12'$, as shown in Fig.~\ref{fig:A2744_xis}. The
azimuthal angles in three directions are NW: $335^{\circ}-90^{\circ}$,
NE: $90^{\circ}-220^{\circ}$, and S: $220^{\circ}-335^{\circ}$,
respectively, and All: $0^{\circ}-360^{\circ}$ is used for the
azimuthally averaged profile. For the NE direction, $8'-12'$ is
  outside the XIS field of view. No significant cluster emission is
detected beyond $12\arcmin$ from the cluster center in the S
direction, which was not used in the present analysis.

\begin{figure}[htbp]
\centering
\rotatebox{0}{\scalebox{0.33}{\includegraphics{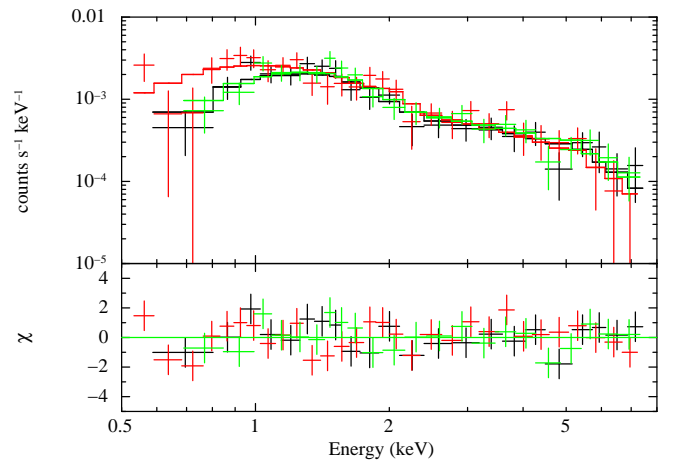}}}
\caption{Example of analysis of annular spectra of
  \S\ref{subsec:annular_spec}. Fit to XIS spectra in NE
  $5\arcmin-8\arcmin$ region is based on APEC model.}
   \label{fig:ne2_fit}%
 \end{figure}

\begin{figure}[htbp]
\centering
\rotatebox{0}{\scalebox{0.33}{\includegraphics{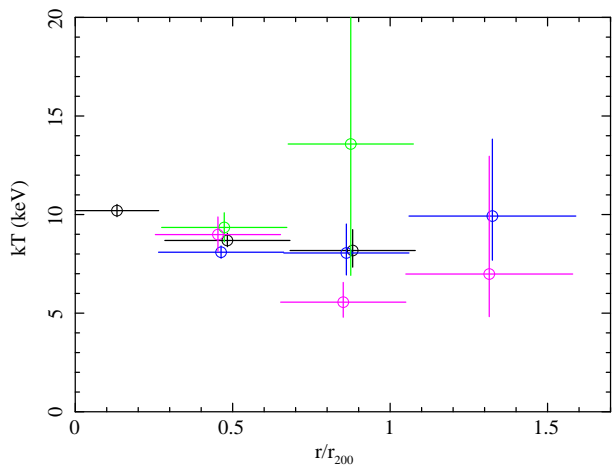}}}
\caption{Projected temperature profiles obtained from XIS spectral
  analysis: All (black), NW (blue), NE (green), S (magenta). }
   \label{fig:temp_projected}%
 \end{figure}

XIS-0, XIS-1, XIS-3 spectra in each ring were then simultaneously fit
to the APEC model in the same manner as mentioned above.  
  Figure~\ref{fig:ne2_fit} shows an example of spectral fitting for
  the case of the NE region.  The best-fit parameters are listed in
Table~\ref{tab:annular_spec} and the projected temperature profiles
are plotted in Fig. \ref{fig:temp_projected}.

 We successfully obtained the radial temperature profile up to
 $r_{200}$ for NE and $1.5 r_{200}$ for NW and S for the first time.
 The temperature near the virial radius is associated with relatively
 large statistical uncertainty, however, the result suggests that the
 profile is rather flat.  The temperature outside $r_{200}$ is
 comparable to the global temperature ($\sim 9$~keV) and does not
 decline significantly, as is often seen in other clusters.

 Furthermore, the temperature distribution is anisotropic as there is
 also a hint of temperature jump near $r_{200}$ in the NE while the
 temperature is lowest in the S direction at the same radius, which
 will be discussed in more detail later.

\begin{table*}[htb]
\centering
\caption{APEC model parameters obtained from analysis of annular spectra from \object{A2744}}\label{tab:annular_spec}
\begin{tabular}{lllll} \hline\hline
 \multicolumn{5}{c}{All} \\ \hline
  Region & $0' - 2'$$^{\mathrm{a}}$ & $2' - 5'$ & $5' - 8'$  & --\\ \hline
  $kT$~[keV] & $10.20^{+0.31}_{-0.31}$ & $8.69^{+0.36}_{-0.32}$& $8.16^{+1.05}_{-0.84}$ & --\\ 
  $Z$~[solar] & $0.24^{+0.04}_{-0.04}$ & $0.19^{+0.04}_{-0.04}$ & $0.24^{+0.13}_{-0.12}$&--\\ 
  Norm & $5.49^{+0.81}_{-0.81}\times10^{-3}$ & $4.66^{+0.07}_{-0.08}\times10^{-3}$ & $1.24^{+0.06}_{-0.06}\times10^{-3}$ & --\\ 
  $\chi^{2}/{\rm d.o.f.}$ & 1032/956 & 841/750 & 323/289  &\\ \hline\hline
\multicolumn{5}{c}{NW} \\ \hline
 Region & $0' - 2' $& $2' - 5'$ & $5' - 8'$ & $8' - 12'$ \\ \hline
 $kT$~[keV] & -- & $8.09^{+0.31}_{-0.31}$& $8.05^{+1.47}_{-1.12}$ & $9.92^{+3.90}_{-2.24}$ \\ 
 $Z$~[solar] & -- & $0.23^{+0.05}_{-0.05}$ & $0.17^{+0.16}_{-0.16}$ & $0.38^{+0.43}_{-0.38}$ \\ 
 Norm & -- & $5.13^{+0.11}_{-0.11}\times10^{-3}$ & $1.26^{+0.09}_{-0.08}\times10^{-3}$ & $1.05^{+0.13}_{-0.12}\times10^{-3}$\\ 
 $\chi^{2}/{\rm d.o.f.}$ & -- & 570/546 & 246/197 & 221/189  \\ \hline\hline
\multicolumn{5}{c}{NE} \\ \hline
 Region & $0' - 2' $& $2' - 5'$ & $5' - 8'$ & $8' - 12'$ \\ \hline
 $kT$~[keV] & -- & $9.35^{+0.74}_{-0.73}$ & $13.58^{+8.17}_{-4.41}$ & -- \\ 
 $Z$~[solar] & -- & $0.21^{+0.08}_{-0.08}$ & $0.29^{+0.75}_{-0.29}$ & -- \\ 
 Norm & -- & $3.98^{+0.13}_{-0.13}\times10^{-3}$ & $1.11^{+0.15}_{-0.16}\times10^{-3}$ & --\\ 
 $\chi^{2}/{\rm d.o.f.}$ & -- & 283/294 & 60.6/73 & - \\ \hline\hline
\multicolumn{5}{c}{S} \\ \hline
 Region & $0' - 2' $& $2' - 5'$ & $5' - 8'$ & $8' - 12'$ \\ \hline
 $kT$~[keV] & -- & $8.98^{+0.90}_{-0.85}$ & $5.56^{+1.00}_{-0.77}$ &$6.99^{+5.97}_{-2.16}$ \\ 
 $Z$~[solar] & -- &$0.06^{+0.09}_{-0.06}$ & $0.25^{+0.23}_{-0.21}$ &$0.61^{+0.90}_{-0.61}$ \\ 
 Norm & -- & $4.56^{+0.17}_{-0.16}\times10^{-3}$ & $1.32^{+0.13}_{-0.13}\times10^{-3}$ & $0.63^{+0.17}_{-0.16}\times10^{-3}$ \\ 
 $\chi^{2}/{\rm d.o.f.}$ & -- & 271/236 & 170/145 & 45.3/62 \\ \hline
\end{tabular}
\begin{list}{}{}
\item[$^{\mathrm{(a)}}$] The inner-most region is a circle with $r=2'$, which is common among the three directions. 
\end{list}
\end{table*}

\subsection{Systematic errors in temperature measurement}\label{subsec:syserr}
To check the robustness of the temperature measurement in the cluster
outskirts, we consider possible systematic errors due to (i) the
background model and (ii) the effect of the XRT's point spread
function (PSF), and we examine (iii) possible contamination of the
cluster spectra from point sources in the NE region.

For (i), we found that the 2--10~keV CXB flux estimated from A2744
south is 30\% lower than that of a blank-sky spectrum from the XIS
data of the Lockman Hole and 36\% lower than that obtained with the
{\it ASCA} Medium Sensitive Survey and Large Sky Survey
\citep{2002PASJ...54..327K}. We confirmed this trend by comparing the
{\it ROSAT} PSPC spectra of annular regions around A2744 and around
the Lockman Hole that we retrieved from the {\it ROSAT} All-Sky Survey
diffuse background map \citep{1997ApJ...485..125S}.  Considering the
typical large-scale fluctuations of $1\sigma\sim 7$\%
\citep{2002PASJ...54..327K}, the probability that we would observe
this low intensity is likely to be low, however, we use the present
CXB model in Table~\ref{tab:bgd} as the nominal case for A2744 and
examine the impact of background uncertainty in a more quantitative
manner below.

The statistical fluctuation of the source number count in the XIS
field of view gives the CXB brightness fluctuation expressed as
$\sigma_{\rm CXB}/I_{\rm CXB}\propto \Omega_{\rm Bgd}^{-0.5}S_{\rm
  c}^{0.25}$. Here $\Omega_{\rm Bgd}$ is the effective solid angle and
$S_{\rm c}$ is the energy flux of the faintest point source detectable
in the field of view.  Following the same approach as in
\citet{2010PASJ...62..371H}, we estimate the CXB fluctuation by
substituting into the above relation $\Omega_{\rm Bgd} = 0.0069~{\rm
  deg^{2}}$ and $S_{\rm c}=1.0\times10^{-14}~{\rm
  erg\,s^{-1}\,cm^{-2}}$ from the present XIS observation. With this
substitution and referring to the values $\sigma_{\rm CXB}/I_{\rm
  CXB}=5\%$ and $S_{\rm c}=6\times10^{-12}~{\rm erg\,s^{-1}\,cm^{-2}}$
from the {\it Ginga} satellite \citep{1989PASJ...41..373H}, we obtain
$\sigma_{\rm CXB}/I_{\rm CXB}=13$\%. We thus intentionally change the
background intensity by $\pm 13$\% and find that this does not
significantly influence the resulting spectral parameters.

For (ii), the {\it Suzaku} XRT has a wide PSF (half power diameter
$2\arcmin$) that allows photons from adjacent regions to contaminate
the data.  To avoid significant photon mixing, we set the width of
annular regions larger than $2\arcmin$ for the outer
rings. Simulations by {\it xissim} raytracing show that, in terms of
the fraction of photons, the PSF effect is 40\% for $r=5'-8'$ whereas
it is as small as 20\% for the outer-most ($r=8'-12'$) ring.

(iii) In the NE region, many point sources were detected by {\it
  XMM-Newton} (Fig.~\ref{fig:A2744_xis}), and the contamination of the
cluster spectra from their emission may not be negligible, although
they were removed with the $r=1\arcmin$ circle. To check their impact,
we extracted {\it XMM-Newton} spectra of the point sources and fitted
them to the power-law model. Because of the poor photon statistics
(only 260 counts), the photon index was fixed at $\Gamma=1.5$. The
resulting power-law flux was $(1.3 \pm 0.2) \times 10^{-13}~{\rm
  erg\,s^{-1}cm^{-2}}$ in the 0.2--12~keV band, which corresponds to
30\% of the observed flux in the region of interest. Then, we fitted
the XIS spectra of the entire NE $5'-8'$ region to a two-component
model consisting of the APEC model and the $\Gamma=1.5$ power-law
model whose normalization was fixed to the best-fit value. The gas
temperature was obtained as $kT=8.2^{+2.3}_{-1.5}$~keV in this
case. This is marginally lower than the value in
Table~\ref{tab:annular_spec}, $kT=13.6^{+8.2}_{-4.4}$~keV but has an
overlap within the errors. To accurately estimate the errors, we
assigned $\Delta T = 5$~keV (a difference between the two best-fit
values) to a systematic error of the temperature in NE $5'-8'$
region. By adding statistical and systematic errors in quadrature, we
obtained $kT=13.6^{+9.6}_{-6.7}$~keV. Note that in
Figs.~\ref{fig:temp_projected}--\ref{fig:comp_temp} and
Eq.~\ref{eq:mass_ne}, we showed error ranges calculated by considering
systematic errors.

\subsection{Deprojection analysis}\label{subsec:deproj}
To derive the radial profiles of gas density and entropy, we performed
a deprojection analysis of the annular spectra under the assumption
that the cluster gas distribution is spherically symmetric. Although
we know that A2744 has an irregular X-ray morphology, the deprojection
analysis allows the gas properties to be compared with those of other
clusters (deduced under the same assumption).

\begin{figure*}[hbt]
   \centering
\rotatebox{0}{\scalebox{0.33}{\includegraphics{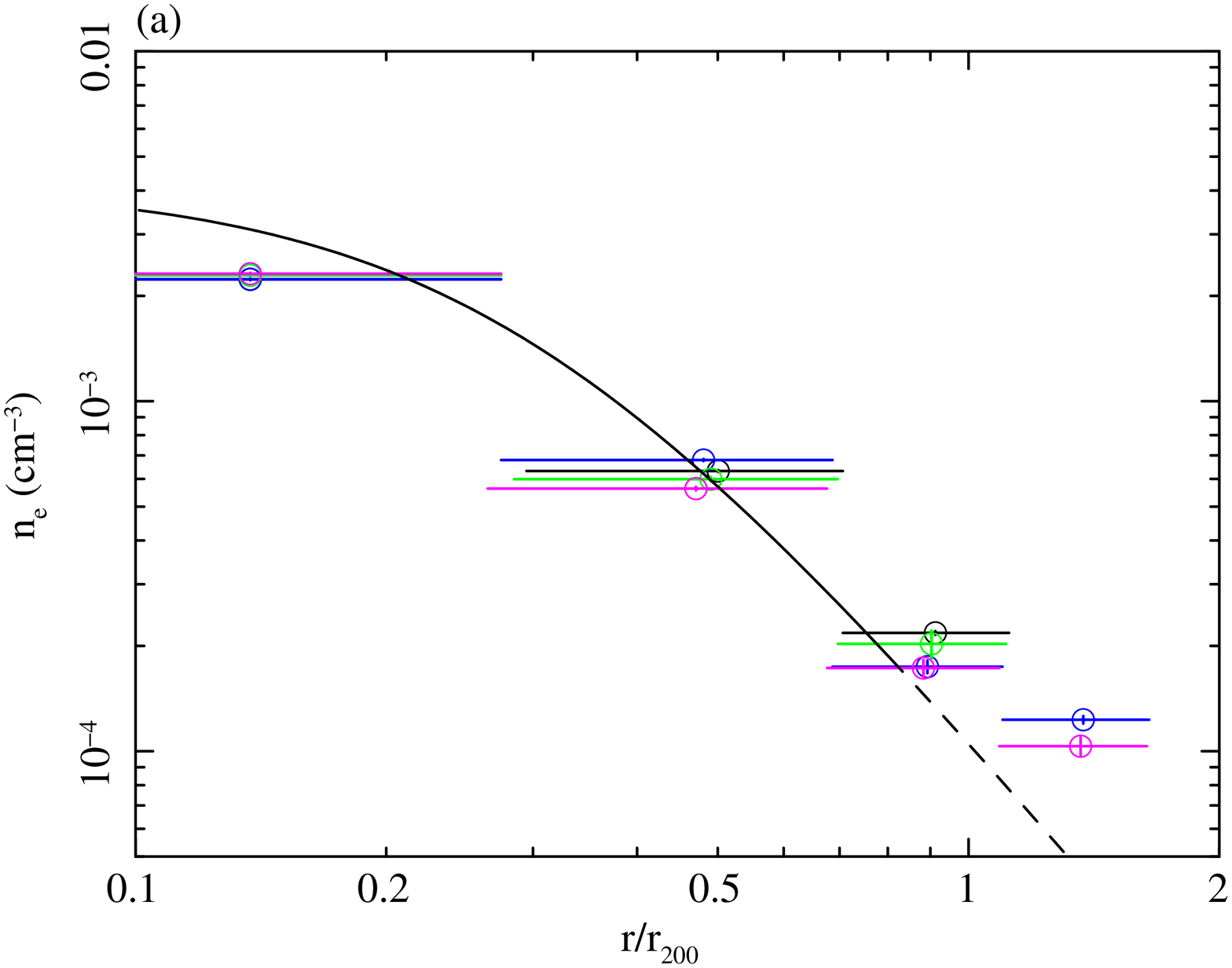}}}
\rotatebox{0}{\scalebox{0.33}{\includegraphics{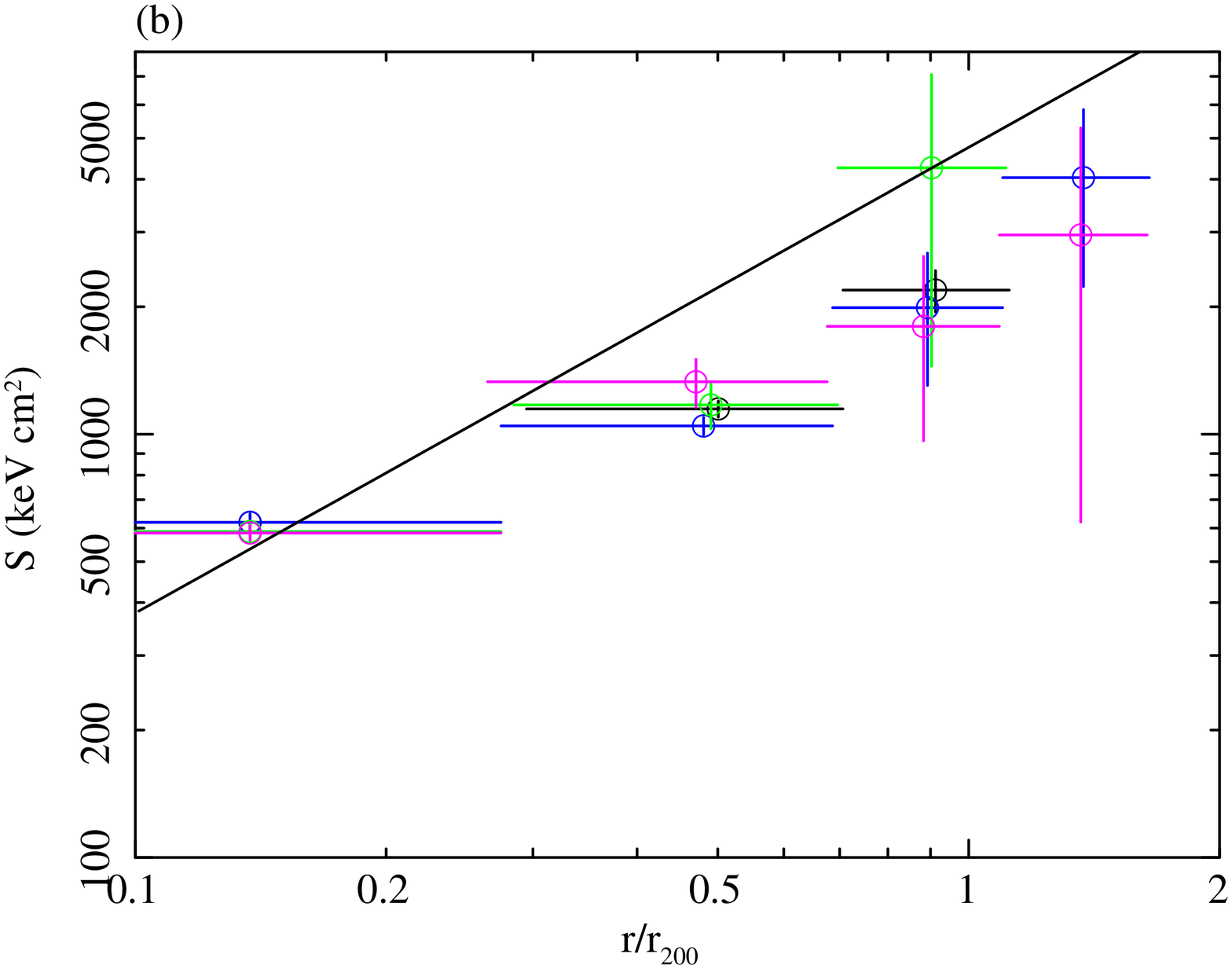}}}
\caption{(a) Electron density and (b) entropy profiles derived from
  deprojection analysis. The meanings of colors are the same as
    in Fig.~\ref{fig:temp_projected}. In panel (a), the solid line
  represents the $\beta$-model from {\it ROSAT} data
  \citep{2004A&A...428..757O}.  Since the $\beta$-model parameters
  were derived within $r<1.6$~Mpc, its extrapolation outside that
  radius is shown by the dashed line.  In panel (b), the solid line
  shows the baseline entropy profile, $S \propto r^{1.1}$
  \citep{2005MNRAS.364..909V}.  }
     \label{fig:depro}%
 \end{figure*}

The APEC model corrected for the Galactic absorption was fit to each
radial bin with the assumption that the metal abundance is radially
constant.  The arithmetic deprojection operation was performed by the
``projct'' model in {\tt XSPEC}.  The two panels in
Fig.~\ref{fig:depro} show the electron density and entropy profiles
where the radius is normalized by $r_{200}$. The electron density was
calculated from the APEC normalization factor $\int n_e n_H
dV/(4\pi(1+z)^2 D_A^2)~[10^{-14}~{\rm cm^{-5}}]$, where $n_e = 1.2
n_{\rm H}$. The entropy, defined as $S\equiv kT n_e^{-2/3}$
\citep{2005RvMP...77..207V} is calculated by substituting the electron
density and the temperature derived from the deprojection analysis for
each radial bin.
 
\section{Discussion}\label{sec:discussion}
With {\it Suzaku}, the profiles of temperature, gas density, and
entropy were measured out to $r_{200}\sim2$~Mpc in A2744.  In this
section, to discuss the gas heating process in the cluster outskirts
and the evolutionary stage of this unique object, the gas properties
of \object{A2744} are compared with X-ray observations of other
clusters as well as with multi-wavelength data of this cluster.

\subsection{Gas properties in cluster outskirts}
\begin{figure}[htbp]
\centering
\rotatebox{0}{\scalebox{0.33}{\includegraphics{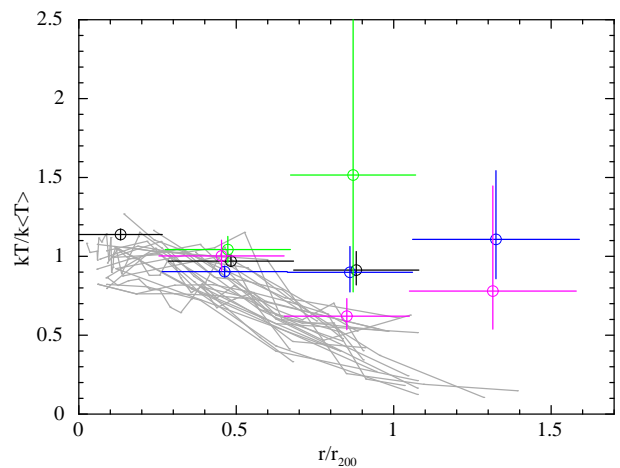}}}
\caption{Comparison of radial temperature profile between A2744 
    All (black), NW (blue), NE (green), S (magenta) and {\it Suzaku}
  sample \citep[gray lines,][]{2013SSRv..tmp...56R}. }
   \label{fig:comp_temp}%
 \end{figure}

Figure~\ref{fig:comp_temp} compares \object{A2744}'s
azimuthally-averaged temperature profile (\S\ref{subsec:annular_spec})
with currently available {\it Suzaku} profiles taken from the
literature \citep[][and references
therein]{2013SSRv..tmp...56R}. Although the nearby clusters show a
systematic temperature decline by a factor of 3 to $r_{200}$, A2744
has a flat temperature distribution and the temperature near $r_{200}$
is one of the highest among them. This result indicates that some
significant heating process is at work in the cluster outskirts.

 In Fig.~\ref{fig:depro} (a), the best-fit $\beta$-model gas-density
 profile obtained from the {\it ROSAT}/HRI data
 \citep{2004A&A...428..757O} is shown for comparison. Because the
 $\beta$-model analysis was performed by using the cluster image
 within $r<1.6$~Mpc, the density profile is extrapolated outside that
 radius. Overall, the present result agrees well with that of the
 $\beta$-model within the virial radius for all directions. On the
 other hand, some flattening is seen at the outermost region. A
 similar effect was reported for other clusters such as A1689
 \citep{2010ApJ...714..423K} and A1795 \citep{2009PASJ...61.1117B}.

 Based on simulations considering gravitational heating because of
 smooth mass accretion \cite{2005MNRAS.364..909V} noted that the
 radial gas entropy profile in clusters tends to follow a power-law,
 $S(r) \propto T_{200} r^{1.1}$ that can be used as a baseline for
 assessing the impact of non-gravitational processes in the
 ICM. Figure~\ref{fig:depro}(b) shows the baseline profile calculated
 for $kT_{200}=9.0$~keV. The entropy profiles at $r<r_{200}$ are
 flatter than that predicted by the baseline profile, as observed in
 other nearby clusters \citep{2012MNRAS.427L..45W}. At $r\gtrsim
 r_{200}$, however, the entropy of A2744 shows no clear drop in the
 outskirts, unlike the previous {\it Suzaku} results
 \citep{2012MNRAS.427L..45W}. This suggests that the thermal
 properties of ICM in the cluster outskirts should be significantly
 affected by the complex merging activities in this system, which are
 examined below by comparing them with multi-wavelength observations.

\subsection{Comparison between X-ray and optical observations}
To examine a relationship between the gas property and galaxy
distribution in outer regions of the cluster, we compare the gas
temperature with the galaxy surface density. The galaxy density was
calculated by using the galaxy catalogue in Table~5 of
\cite{2011ApJ...728...27O}, who compiled two photometric catalogs
based on the Supercosmos Sky Survey \citep{2002A&A...389..787B} and
the Very Large Telescope \citep{2009A&A...500..947B}. We selected 442
galaxies that meet the same criterion as used in
\citet{2006A&A...449..461B}: $cz_{\rm cluster} \pm 5000~{\rm
  km\,s^{-1}}$ (or $z_{\rm cluster} \pm 0.0167$) and the $r_F$-band
magnitude $r_{\rm F}<23$.  The following result does not significantly
change if we select galaxies with the same criteria as used in A1835
at $z=0.25$, $r_{\rm F} < 22$ and $cz _{\rm cluster} \pm 3000~{\rm
  km\,s^{-1}}$ \citep{2013ApJ...766...90I}.
 
\begin{figure}[htb]
\centering
\rotatebox{0}{\scalebox{0.33}{\includegraphics{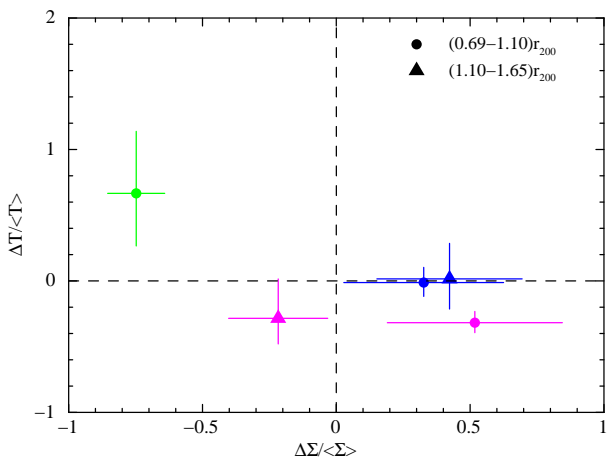}}}
\caption{Comparison of temperature deviation $\Delta T/<T>$ and galaxy
  surface density contrast $\Delta \Sigma/<\Sigma>$ in outskirts of
  \object{A2744}. The data for the two radial bins $(0.7-1.1)r_{200}$
  and $(1.1-1.6)r_{200}$ are shown by circles and triangles,
  respectively. The meanings of colors are the same as in
    Fig.~\ref{fig:temp_projected}. The quoted errors are the
  $1\sigma$ statistical uncertainties. }
   \label{fig:xvsopt}%
 \end{figure}

Table~\ref{XvsOpt} shows the gas temperature measured by {\it Suzaku}
and the galaxy density for two radial ranges; namely, $0.69<
r/r_{200}<1.10$ and $1.10< r/r_{200}<1.65$. The ranges are normalized
by their mean values in each radial bin and the errors indicate the
statistical uncertainties at 68\% confidence. In
Fig.~\ref{fig:xvsopt}, the temperature deviation $(T-<T>)/<T>$ is
plotted as a function of the galaxy density contrast,
$(\Sigma-<\Sigma>)/<\Sigma>$.

As is clear from Table~\ref{XvsOpt}, the galaxy density is
anisotropic. For $0.69< r/r_{200}<1.10$, we confirm the galaxy excess
in the NW and S directions where the filaments have already been
identified by \citet{2007A&A...470..425B}. On the other hand, the
galaxy density is significantly lower in the NE direction. For larger
radii ($1.10< r/r_{200}<1.65$), however, the density in the NE
direction is comparable to the mean. If compared with the gas
temperature (Fig.~\ref{fig:xvsopt}), no simple one-to-one
correspondence exists between the optical and X-ray properties.  These
results differ from previous reports on \object{A1689} and
\object{A1835} that claimed a positive correlation between galaxy
density and gas temperature
\citep{2010ApJ...714..423K,2013ApJ...766...90I}. Therefore, the
present result strongly suggests a complex dynamical state and mass
structure in A2744.

\begin{table*}[htb]
	\centering
	\caption{Gas temperature and galaxy surface density in outer regions of A2744}\label{XvsOpt}
	\begin{tabular}{lllll} \hline\hline
	 &\multicolumn{2}{c}{$r/r_{200}=0.69-1.10$} &\multicolumn{2}{c}{$r/r_{200}=1.10-1.65$}\\ \hline
	 Region & X-ray temperature &Galaxy surface density & X-ray temperature &Galaxy surface density \\ 
	  & $kT$~[keV] & $\Sigma~[{\rm arcmin^{-2}}]$ & $kT$~[keV] & $\Sigma~[{\rm arcmin^{-2}}]$ \\ \hline 
	  NW        & $8.05^{+0.78}_{-0.68}$  & $0.72 \pm 0.14$ &$9.92^{+2.10}_{-1.56}$  &$0.50 \pm 0.08$\\ 
          NE         & $13.58^{+3.74}_{-3.14}$ & $0.14 \pm 0.06$ &--$^{\mathrm{a}}$          & $0.29 \pm 0.06$ \\ 
 	  S           & $5.56^{+0.60}_{-0.50}$   & $0.82 \pm 0.15$ &$6.99^{+2.70}_{-1.53}$ & $0.27 \pm 0.06$\\  \hline
	  Average & $8.15\pm 0.53$              & $0.54 \pm 0.07$ & $9.77\pm1.58$             & $0.35 \pm 0.04$ \\ \hline
	\end{tabular}
\begin{list}{}{}
\item[] The $1\sigma$ statistical errors are quoted. 
\item[$^{\mathrm{(a)}}$] For NE, $r/r_{200}=1.10-1.65$ is outside the XIS field of view. 
\end{list}
	\end{table*}

\subsection{Comparison between X-ray and radio observations}
\begin{figure}[htb]
\centering
\rotatebox{0}{\scalebox{0.4}{\includegraphics{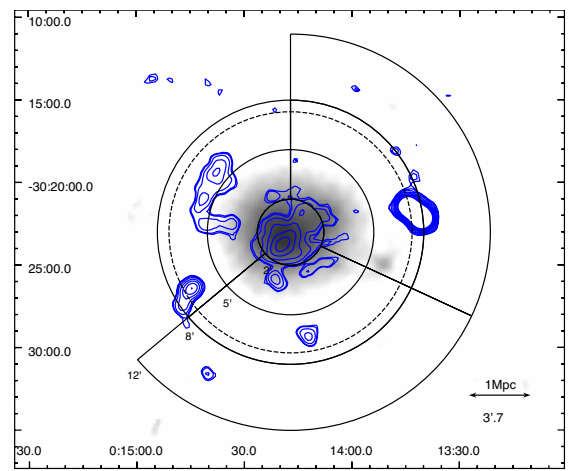}}}
\caption{{\it Suzaku}-XIS image (gray scale) overlaid with radio
  325~MHz intensity contours (blue) \citep{2007A&A...467..943O}. A
  radio halo and radio relic extending over 1~Mpc in length exist at
  the cluster core and at $7\arcmin$ (or $0.8r_{200}$) north-east of
  the center, respectively. The XIS spectral regions (black) and
  virial radius ($r_{200}=7'.3$) are also indicated.}
   \label{fig:radio}%
 \end{figure}

From the {\it Suzaku} observation, we found a hint of temperature jump
in the NE region. In fact, as seen from Fig.~\ref{fig:radio}, the
location of the high-temperature region coincides well with that of
the large radio relic. Thus, the gas may have undergone shock
  heating because of merging or mass accretion onto the main cluster.
For \object{A3667}, a sharp change of both temperature and brightness
near the relic was observed
\citep{2010ApJ...715.1143F,2012PASJ...64...49A}. With the current
photon statistics of the {\it Suzaku} data in the NE region, however,
we find no significant discontinuity in the X-ray surface brightness
at the radio relic. To clarify the origin of the high-temperature gas
in the NE region and its relationship to the radio relic, we will
examine the temperature and density structures in more detail upon
completion of an additional pointing observation approved for {\it
  Suzaku} AO8.

\subsection{X-ray mass estimation and implications}
The complex gas distribution in A2744 indicates that the cluster is
not in the relaxed state. However, we expect that comparing the
hydrostatic mass estimated from X-ray observations with the
gravitational lensing mass will clarify the dynamical state of the
system
\citep[e.g.,][]{1998ApJ...495..170O}. \cite{2011MNRAS.417..333M}
estimated the weak lensing mass of this cluster to be $M_{\rm
  lens}(r<1.3~{\rm Mpc})=(1.8 \pm 0.4)\times10^{15}M_{\sun}$. From the
radially-averaged temperature profile (All:$0-360^{\circ}$) derived in
\S\ref{subsec:deproj} and the $\beta$-model parameters $\beta=0.96$
and $r_c=133$~kpc \citep{2004A&A...428..757O}, we calculate the
hydrostatic mass projected within a radius of 1.3~Mpc to be
\begin{equation}
M_{\rm X} (<1.3~{\rm Mpc}) = 1.7^{+0.2}_{-0.2}\times 10^{15}~{\rm M_{\sun}}.
\end{equation} 
If the temperature profiles in three different directions are separately used, we obtain
\begin{eqnarray}
M_{\rm X} (<1.3~{\rm Mpc}) &=& 1.4^{+0.5}_{-0.4}\times 10^{15}~{\rm M_{\sun}}~~{\rm for~NW}, \\
M_{\rm X} (<1.3~{\rm Mpc}) &=& 2.9^{+2.6}_{-1.9}\times 10^{15}~{\rm M_{\sun}}~~{\rm for~NE}, \label{eq:mass_ne}\\
M_{\rm X} (<1.3~{\rm Mpc}) &=& 1.3^{+0.7}_{-0.4}\times 10^{15}~{\rm M_{\sun}}~~{\rm for~S}.
\end{eqnarray}
Excluding the NE region where the high-temperature gas exists, the
systematic uncertainty in the X-ray hydrostatic mass is estimated to
be 25\%. Although the X-ray mass derived from the mean temperature
profile is in an agreement with the weak lensing mass within the
statistical error, its interpretation must be done with care given the
systematic uncertainty.

Based on X-ray observations \citep[e.g.,][]{2012PASJ...64...49A,
  2012PASJ...64...67A} and hydrodynamic simulations
\citep[e.g.,][]{2010PASJ...62..335A}, a possibility of non-equilibrium
ionization in merging clusters has been pointed out. Assuming that the
ionization timescale is given by $n_e t =10^{12}~{\rm cm^{-3}\,s}$
\citep{1984Ap&SS..98..367M} and the observed high-temperature gas
originates from the merger shock having a shock velocity of the order
$2000~{\rm km\,s^{-1}}$, the ion-electron relaxation timescale near
the viral radius of A2744 is likely to be longer than the time elapsed
after the shock occurs. A more quantitative evaluation of the physical
state of the ICM requires an additional observation.

\section{Summary}\label{sec:summary}
By using the {\it Suzaku} XIS detectors, we performed an X-ray
spectral analysis of the merging cluster \object{A2744} at $z=0.3$ and
derived the temperature profiles out to large radii ($r_{200}$ in the
NE direction and $1.5 r_{200}$ in the NW and S directions) for the
first time. The temperature is as high as $kT\sim 9$~keV even near the
virial radius and does not decline significantly in the outskirts,
which differs from all other clusters observed with {\it Suzaku}. We
also found that the temperature structure is anisotropic and exhibits
no clear positive correlation with the galaxy surface density,
suggesting that the cluster has a very complex mass structure and is
dynamically young. We find a hint of temperature jump in the NE region
whose location coincides well with that of a large radio relic. Thus
the gas may have undergone shock heating because of merging or mass
accretion.  A further examination of the shock structure near the
radio relic must await a follow-up X-ray observation of the NE region.


\begin{acknowledgements}
  We are grateful to the {\it Suzaku} team members for satellite
  operation and instrumental calibration.  We would also like to thank
  E. Orr\`u for providing the radio data, T. H. Reiprich for the
  electronic data to reproduce a figure of temperature profiles
  obtained with {\it Suzaku}, and D. Pierini and A. Scaife for useful
  discussions.  We are also grateful to K. Sato for suggesting a
  useful means of calculating the PSF effect. This work is in part
  supported by a Grant-in-Aid by MEXT, KAKENHI Grant Number
  25400231(NO).  Y.Y.Z. acknowledges support by the German BMWi
  through the Verbundforschung under grant 50\,OR\,1103.
\end{acknowledgements}

\bibliographystyle{aa}
\bibliography{a2744_aa}

\end{document}